\documentclass[10pt,
preprintnumbers,amsmath,amssymb,
aps,prd,nofootinbib,eqsecnum,a4paper]{revtex4}
\usepackage{graphicx,epsf}
\usepackage{xcolor}
\usepackage{relsize}
\def\be{\begin{equation}}
\def\ee{\end{equation}}
\def\bea{\begin{eqnarray}}
\def\eea{\end{eqnarray}}
\def\p{\partial}

\def\cs2{c_{\rm{s}}^2}
\def\wt{\widetilde}
\def\frw{{\rm{FRW}}}
\def\SMTP{{\zeta}_{\rm{SMTP}}}
\def\DOTSMTP{{\dot{\zeta}}_{\rm{SMTP}}}

\newcommand{\sfx}{X}
\newcommand{\sfy}{Y}
\newcommand{\dpn}{\delta P_{\mathrm{nad}}}
\newcommand\eq[1]{Eq.~(\ref{#1})}
\def\beal{\begin{align}}
\def\eeal{\end{align}}
\def\p{\partial}

\begin{document}

\title{Conserved Quantities in Expanding G{\"o}del Cosmology}
\author{Alexander Leithes}

\affiliation{Astronomy Unit, School of Physics and Astronomy,
Queen Mary University of London, Mile End Road, London, E1 4NS, UK}
\affiliation{The Roche School, 11 Frogmore, London, SW18 1HW, UK}

\date{\today}

\begin{abstract}
At linear order we study perturbations to a G{\"o}del background spacetime which includes expansion in addition to rotation. We investigate the transformation behaviour of these
perturbations under gauge transformations and construct gauge
invariant quantities. Using the perturbed energy conservation
equation we find that there are conserved quantities in Expanding G{\"o}del (EG) Cosmology, in particular a spatial metric trace perturbation, $\SMTP$, which is conserved on large scales for pressureless dust. We finally extend our discussion to a perfect fluid matter content to also obtain conserved quantities in this context.
\end{abstract}

\maketitle

\section{Introduction}
\label{Introduction}

Recent discoveries by the James Webb Space Telescope (JWST) of an apparent bias in the alignment of the axes of rotation of galaxies throughout the observable universe ~\cite{JWST} (specifically that 50\% more galaxies surveyed rotate in the opposite direction relative to the Milky Way galaxy compared to those rotating in the same direction) has added weight to the earlier evidence of the same phenomenon discovered from, for example, the SDSS data sets ~\cite{SDSS}.
One mechanism which has been proposed as driving this alignment is that of a rotating cosmology ~\cite{JWST}. Specifically in ~\cite{JWST} it is suggested that a cosmological-scale axis of alignment for galaxy rotations implies that the entire cosmology may intrinsically have such an axis, as would be the case in a rotating cosmology.\\
A rotating cosmology has also been suggested as possible method of resolving the Hubble tension between early and late time derived values for the Hubble constant ~\cite{HUBBLETENSION}.\\

The earliest and perhaps best known of these cosmologies is that proposed by Kurt G{\"o}del in 1949 ~\cite{GODEL1949}, which famously gave exact solutions to the Einstein Field Equations. 
While this cosmology includes rotation it does not allow for expansion of the universe as supported by current observations e.g. ~\cite{Hubble}.
In 2000 Yuri N. Obukhov constructed an extension to G{\"o}del's cosmology which does allow for an expanding as well as rotating universe ~\cite{REFEG}, in which the author states it is an exact solution as per the original G{\"o}del cosmology. Hereafter this extension is referred to as Expanding G{\"o}del (EG) cosmology.
Indeed, ~\cite{HUBBLETENSION} specifically cites ~\cite{REFEG} as the model they are likely to use in further studies in the context of rotating cosmologies.\\

In order to study structure formation in such a cosmology it would be helpful to discover conserved quantities in that spacetime. More specifically, in cosmological perturbation theory (see e.g. \cite{Bardeen80,KS}), these gauge invariant conserved perturbations enable us to relate early to late times in a given cosmological model (e.g.~\cite{Lyth85}), the standard cosmological model usually being a Friedmann-Robertson-Walker (FRW) \cite{Friedmann} background spacetime.\\

In this work we build on both the previous work in gauge invariant conserved quantities in FRW, for example the curvature perturabtion $\zeta$ , as well as our previous work in Lema{\^\i}tre-Tolman-Bondi (LTB) \cite{Bondi} and Lema{\^\i}tre spacetimes ~\cite{Leithes2014} where we construct a similar conserved quantity $\SMTP$ which was conserved in those spacetimes as well as in FRW.
To avoid confusion we point out at this stage that FRW is the standard model used in cosmology today and has the background metric,
\be 
\label{ds2}
ds^2=-dt^2+a(t) ^2\delta_{ij}dx^idx^j \,, 
\ee
where $a (t)$ is the time only dependent scale factor.
LTB spacetime has the following background metric,
\begin{equation}
\label{LTBInterval}
ds^2 = -dt^2 + \sfx^2 (r,t) dr^2 
+ \sfy^2 (r,t) \left( d \theta^2 + \sin^2 \theta d \phi^2 \right) ,
\end{equation}
where $\sfx (r,t)$ and $\sfy (r,t)$ are scale factors dependent upon both the
radial spatial and time co-ordinates and are not independent. 
Finally, Lema{\^\i}tre spacetime differs from LTB spacetime and has the following background metric,
\begin{equation}
\label{LInterval}
ds^2 = -f^2 dt^2 + \sfx^2 (r,t) dr^2 + \sfy^2 (r,t) \left( d \theta^2 + \sin^2 \theta d \phi^2 \right) \,,
\end{equation}
where $f$ is an additional third scale factor, $f \equiv f(r,t)$.
These metrics are only shown here for reference and for comparison with the EG metric to follow. For a more detailed analysis of these other cosmologies see e.g.~\cite{Leithes2014}.
\\

It should be noted that objections have been made in the past to the compatibility of rotating cosmologies with an expanding universe more generally. Somewhat analogous to a pirouetting ice-skater throwing out their arms to slow their rotation through conservation of angular momentum, it has been suggested that in an expanding universe the effects of any rotation would rapidly become negligible and undetectable throughout most of the history of the universe. Indeed, Obukhov briefly discusses these objections in his paper \cite{REFEG}.
However, such a diminishing influence of rotation with time on structure formation seems to be supported by recent observations. The discovery of the alignment of galactic rotations in later epochs measured using SDSS data ~\cite{SDSS} versus the same phenomenon detected to a greater degree through earlier epochs using JWST data ~\cite{JWST} might be expected. In later times the alignment may have become disrupted to some degree by the peculiar motions of individual galaxies and clusters of galaxies and their interactions in the absence of a significant restoring force provided by any residual cosmological rotation. Whereas in earlier times, where the rotation and its aligning influence was more pronounced, the galaxy rotation alignment is correspondingly more significant.
\\

This paper is structured as follows. In the following section \ref{Gov_sect} we present the general governing equations under General Relativity independent of a given spacetime.
In Section \ref{EGBack_sect} we review Expanding G{\"o}del cosmology at the background level for a pressureless dust matter content.  We then extend this by adding perturbations to the Expanding G{\"o}del background. We also examine the transformation behaviour of matter and metric variables and then construct gauge invariant quantities. In the last part of this section we look at the evolution of the spatial metric trace perturbation, $\SMTP$.  Finally in Section \ref{Conclusion} we discuss the implications of our work for studies of Expanding G{\"o}del spacetime, the evolution of structure therein, as well as other areas of further research in Cosmology more widely in which this work may prove useful.

\section{General Governing equations}
\label{Gov_sect}

In this section we discuss the general governing equations from Einstein's General Relativity without assuming any particular spacetime. We also do not yet split quantities into background plus perturbations. Here we restrict the discussion to just those quantities which were useful in this paper.\\

The Einstein field equations take the form
\begin{equation}
\label{EinsteinGen}
G_{\mu \nu} = 8 \pi G \, T_{\mu \nu}\,,
\end{equation}
$G_{\mu \nu}$ being the Einstein tensor, $T_{\mu \nu}$ the
stress-energy-momentum tensor, and $G$ is Newton's constant.
The stress-energy-momentum tensor for a perfect fluid is given by,
\begin{equation}
\label{Tmunu_up}
T^{\mu \nu} = (\rho + P) u^{\mu} u^{\nu} + P\ g^{\mu \nu} \,,
\end{equation}
where $g^{\mu \nu}$ is the metric tensor, $\rho$ is the energy density,
$P$ the pressure, and $u^{\mu}$ the 4-velocity of the fluid.
However, we initially use the stress-energy-momentum tensor for pressureless dust
\begin{equation}
\label{Tmunu_upPless}
T^{\mu \nu} = \rho u^{\mu} u^{\nu} \,,
\end{equation}
The metric tensor for any spacetime is subject to this constraint,
\begin{equation}
\label{MetricConstraint}
g^{\mu \nu}g_{\nu \gamma} = {\delta^{\mu}}_{\gamma} ,
\end{equation}
${\delta^{\mu}}_{\gamma}$ being the Kronecker delta.
We define the 4-velocity as,
\begin{equation}
\label{4vel}
u^{\mu} = \frac{dx^{\mu}}{d\tau}\,,
\end{equation}
$\tau$ being the proper time along curves to which $u^\mu$ is tangent, related to the line element $ds$ through
\begin{equation}
\label{Propertime}
ds^2 = - d \tau^2 \,.
\end{equation}
The constraint for the 4-velocity is,
\begin{equation}
\label{4velcons}
u^{\mu}u_{\mu} = -1 \,.
\end{equation}
Energy-momentum conservation is obtained from the Einstein equations,
\eq{EinsteinGen}, via the Bianchi identities,
\begin{equation}
\label{Bianchi}
\nabla_{\mu} T^{\mu \nu} = 0 \,.
\end{equation}
The covariant derivative of a 4-vector can be decomposed as (see e.g. ~\cite{Ellis},~\cite{Wald84}),
\begin{equation}
\label{4vdecomp}
\nabla_\mu u_\nu = -u_\mu u^\alpha \nabla_\alpha u_\nu 
+ \frac{1}{3} \Theta {\cal{P}}_{\mu \nu} 
+ \sigma_{\mu \nu} + \omega_{\mu \nu} \,,
\end{equation}
where we use vector, $u^\mu$, as an example, since \eq{4vdecomp} is true for all 4-vectors including the 4-velocity. Here
$\Theta$ is the expansion factor, $\sigma_{\mu \nu}$ the shear
tensor, $\omega_{\mu \nu}$ the vorticity tensor, 
and ${\cal{P}}_{\mu \nu}$ the spatial projection tensor.

The expansion factor defined with respect to the 4-velocity is, 
\begin{equation}
\label{ExpFacugen}
\Theta = \nabla_{\mu} u^{\mu}\,,
\end{equation}

\section{ Expanding G{\"o}del (EG) spacetime}
\label{EGBack_sect}

Here we first briefly review Expanding G{\"o}del (EG) cosmology at the background level, highlighting minor notation changes we have made from that used by Obukhov in \cite{REFEG}. 
We then extend this by adding perturbations to the EG background. To facilitate removing any unwanted gauge modes, we examine the transformation behaviour of the perturbations, allowing us to construct gauge-invariant quantities, most notably an equivalent to
the curvature perturbation. We then show under which conditions this
curvature perturbation is conserved.

Throughout this section we assume zero background pressure in the
spacetime i.e. a pressureless dust matter content. We do however allow for a pressure perturbation in analysis of the perturbed quantities.
We intend to extend this work to a Perfect Fluid matter content which would include background pressure.

\subsection{Background}
\label{EG_back}

The EG metric as written by Obukov, equation 4.1 in ~\cite{REFEG}, closely follows the original form of the metric in G{\"o}del's non-expanding cosmology ~\cite{GODEL1949} with the addition of a scale factor for the added expansion. We use a slightly modified notation in our version to make it closer to that used in our earlier work in LTB, Lema{\^\i}tre and FRW spacetimes \cite{Leithes2014}, while retaining a similarity to the previous work in FRW more broadly e.g.~\cite{Lyth85}.
In our notation the metric may be written as,
\begin{equation}
\label{EGInterval}
ds^2 = -dt^2 + 2 \sqrt{\sigma} \sfx (x,t) dt dy + a^2 (t) dx^2 + \kappa \sfx^2 (x,t) dy^2 + a^2 (t) dz^2 ,
\end{equation}
where $a (t)$ is the time only dependent scale factor, analogous to that found in standrd FRW cosmology, while $\sfx$ is the scale factor dependent upon both the $x$ spatial and time co-ordinates similar to the scale factors found in LTB cosmology \cite{Leithes2014}. Also similarly to LTB the two scale factors are not independent, but in EG the indices $0, 1, 2, 3$ are $t, x, y, z$ respectively. The scale
factors are related by,
\begin{equation}
\label{ScaleFactorsRelation}
\sfx = a (t) e^{mx} ,
\end{equation}
where $m$ is a constant.
The universal rotation is in the $z$ direction and its magnitude, $\omega$ is given by,
\begin{equation}
\label{omega}
\omega = \frac{m}{2a} \sqrt{\frac{\sigma}{\kappa + \sigma}} ,
\end{equation}
where $m$, $\kappa$ and $\sigma$ are simply constants. It can be seen that in the limit $\sigma=m=0$ and $\kappa=1$ we recover the FRW metric. For a more detailed description of the origin and form of these constants see ~\cite{REFEG}.
\\
It should be noted here that from the form of $\omega$ in \eq{omega} and the term $\frac{m}{2a}$ it can easily be seen that with the rapid expansion of the universe and commensurate increase in the scale factor, $a$, the magnitude of the rotation, $\omega$,  rapidly decreases to negligible levels in late times as expected. Naturally, angular momentum is conserved.
\\
The off-diagonal components of the background metric will clearly lead to more complicated equations in the following sections than those employed in FRW and to some degree even LTB and Lema{\^\i}tre spacetimes.

The background 4-velocity is given from its definition,
\eq{4vel}, as
\begin{equation}
\label{4velunpertLTB}
u^{\mu} = [1,0,0,0] \,,
\end{equation}
as we assume we are co-moving with respect to the background
and therefore, $dx=dy=dz=0$, and hence $d{\tau}^2
= dt^2$ in the local rest frame.

From the definition of the stress-energy-momentum tensor, \eq{Tmunu_up}, we find that for pressureless dust the only non-zero component is, $T^{0 0} = \rho$. For later convenience we define the Hubble parameter and one equivalent parameter for our $X$ scale factor such that,
\begin{equation}
\label{HX}
H = \frac{\dot{a}}{a} \,, \qquad
H_{\sfx} = \frac{\dot{\sfx}}{\sfx} \,.
\end{equation}
where the ``dot'' denotes the derivative with respect to coordinate time $t$.

In order to construct the Einstein equations we must first find the connections, zero and non-zero, denoted by the Christoffel symbols below. In  ~\cite{REFEG} it seems that gauge selections have been made to yield a remarkably small number of non-zero connections - seven including duplicates. Here we make no initial gauge choices choosing to leave our expressions fully general, only later fixing gauges when dealing with perturbations in order to find gauge invariant conserved quantities. As such we find twenty-six non-zero connections as shown in  Appendix \ref{Connections}, with the remaining thirty-eight connections all being zero. Here a ``prime'' denotes the derivative with respect to the spatial coordinate $x$.
\\

In order to derive the Einstein equations we must first construct the components of the Ricci Tensor, $R_{\mu \nu}$, and then the Ricci Scaler, $R$. These may be found in Appendix \ref{RicciTensorCompenents} and Appendix \ref{RicciScalarSec} respectively.
\\

From \eq{EinsteinGen} we find the Einstein equations are as given in Appendix \ref{EFEs}. 
\\

The energy conservation equation in EG spacetime for pressureless dust, obtained from \eq{Bianchi}, is
\begin{equation}
\label{HEGEngConsExpanded}
\dot{\rho} + 2 \frac{\dot{X}}{X}\rho + 2 \frac{\dot{a}}{a}\rho + \frac{\dot{X}}{X}\rho = 0\,.
\end{equation}
We first present here this less concise expanded form of the energy conservation equation rather than the simpler,
\begin{equation}
\label{HEGEngConsContracted}
\dot{\rho} + 3 H_X\rho + 2 H\rho = 0\,,
\end{equation}
as it is not immediately obvious how this could reduce to the energy conservation equation in FRW in the limit $\sigma=m=0$ and $\kappa=1$, i.e.
\begin{equation}
\label{HEGEngConsFRW}
\dot{\rho} + 3 H\rho = 0\,.
\end{equation}
However, the first expression in \eq{HEGEngConsExpanded}, $2 \frac{\dot{X}}{X}\rho$, is actually generated using the off-diagonal terms in the metric \eq{EGInterval}, which means this term vanishes in the limit $\sigma=0$, giving,
\begin{equation}
\label{HEGEngConsfirsttermgone}
\dot{\rho} + 2 \frac{\dot{a}}{a}\rho + \frac{\dot{X}}{X}\rho = 0\,.
\end{equation}
From \eq{ScaleFactorsRelation} it can clearly be seen that in the limit $m=0$ we recover \eq{HEGEngConsFRW} as expected.
Throughout this work we use the Cadabra tensor manipulation package \cite{Cadabra}. While this is an older package it does greatly facilitate the checking of the steps undertaken generating expressions, such as those in \eq{HEGEngConsExpanded}.

\subsection{Perturbations}
\label{pert_EG}

In this section we introduce perturbations to the EG background, while not decomposing these perturbations further into scalar, vector and tensor components. This follows our method in \cite{Leithes2014}, which considerably simplifies our perturbed governing equations.
\\

We do however split quantities into a $t$ and $x$ dependent background part where necessary, and a perturbation depending on all four coordinates. We decompose the energy density $\rho$ as follows,
\be
\label{split_rho}
\rho=\bar\rho(t,x)+\delta\rho(x^\mu)\,,
\ee
here and in the following a ``bar'' denotes a background quantity where there may be a possibility for confusion.
\\

Now we perturb the metric in a similar way as in \cite{Leithes2014} only here we use Cartesian spatial coordinates. We split the metric tensor as, 
\begin{equation}
\label{metricsplit}
g_{\mu \nu} = {\bar{g}}_{\mu \nu} + \delta g_{\mu \nu} , 
\end{equation}
here ${\bar{g}}_{\mu \nu}$ given by \eq{EGInterval}. 
\\

We now make the following ansatz for the perturbed metric, $\delta g_{\mu \nu}$,
\begin{equation}
\label{EGMetricperturbations}
\delta g_{\mu \nu}=\begin{pmatrix}
  -2\Phi & a B_x & \sqrt{\sigma} \sfx B_y & a B_z \\
   a B_x & 2 a^2 C_{xx} & a X C_{xy} & a^2 C_{xz} \\
  \sqrt{\sigma} \sfx B_y & a X C_{xy} & 2 \kappa\sfx^2 C_{yy} & a X C_{yz} \\
  a B_z & a^2 C_{xz} & a X C_{yz} & 2a^2 C_{zz}
\end{pmatrix} \,.
\end{equation}
Here $\Phi$ is the lapse function, while $B_i$, where $i=x,y,z$,
are the shift functions for each spatial coordinate. Similarly,
$C_{ij}$, where $i,j=x,y,z$, form the spatial metric perturbations.\\

From the perturbed metric we can construct the perturbed
4-velocities using the definition, \eq{4vel}.
To linear order Proper time is in the perturbations given by,
\begin{equation}
\label{PropCoordPert}
d{\tau} = (1 +  \Phi) dt \,.
\end{equation}
We define the 3-velocity as, 
\begin{equation}
\label{3veldef}
v^i = \frac{d x^i}{d t} \,.
\end{equation}
From \eq{4vel} we get the contravariant 4-velocity vector,
\begin{equation}
\label{4velpertEG}
u^{\mu} = [(1 - \Phi),v^x,v^y,v^z] .
\end{equation}
By using the perturbed metric we lower the index to obtain the
covariant form of the 4-velocity,
\begin{equation}
\label{4velpertEGDown}
u_{\mu} 
= [-(1 + \Phi) + \sqrt{\sigma} X v^y,
a B_x + a^2 v^x,\,
\sqrt{\sigma} X(B_y + 1 + \Phi) + \kappa X^2 v^y,\,
a B_z + a^2 v^z] \,.\\
\end{equation}
This is more complicated than the equivalent in LTB or Lema{\^\i}tre spacetimes \cite{Leithes2014}, and far more complicated than in FRW (see e.g. \cite{Lyth85}), due in part to the off-diagonal components of the metric even at background \eq{EGInterval}.
\\

As with the background, once again \eq{Bianchi} together with \eq{Tmunu_up} allows us to derive the perturbed energy conservation equation to linear order in perturbations as,
\begin{multline}
\begin{aligned}
\label{PertEconsEG}
& \delta \dot{\rho} + {\bar{\rho}}' v^x + \bar{\rho} {v^x}' + \bar{\rho} \partial_y v^y + \bar{\rho} \partial_z v^z - 2 \dot{\bar{\rho}} \Phi + 2 \dot{B}_y \bar{\rho} + \dot{C}_{xx} \bar{\rho} + \dot{C}_{yy} \bar{\rho} + \dot{C}_{zz} \bar{\rho} + 3 H_X \delta \rho + 2 H \delta \rho + 3 \frac{X'}{X} v^x \bar{\rho} + 2 \frac{\kappa \dot{X}}{\sqrt{\sigma}} v^y \bar{\rho} \\ & - 6 H_X \Phi \bar{\rho} - 4H \Phi \bar{\rho} + \frac{\partial_y \delta P}{\sqrt{\sigma} X} + \frac{\bar{\rho} \partial_y \Phi}{\sqrt{\sigma} X} + \frac{2 \kappa H_X \delta P}{\sigma} + 2H\delta P + H_X \delta P = 0 \,,
\end{aligned}
\end{multline}
where we used \eq{split_rho}, and the pressureless dust background requiring $\bar P=0$. Once again \eq{PertEconsEG} is more lengthy than the equivalents in LTB, Lema{\^\i}tre or FRW.
%

\subsection{Gauge Transformations}
\label{GaugeTransBeh}

As in \cite{Leithes2014} we aim to construct gauge-invariant perturbations, and as such we need to examine the transformation behaviour of our matter and metric variables.  As before, we use the Bardeen approach to cosmological
perturbation theory \cite{Bardeen80, KS}. For an overview with references to the primary literature, see \cite{MW2008}.

Using the active approach, tensorial linear order perturbations, $\mathbf{T}$, transform as
\begin{equation}
\label{gauge_gen}
\delta \tilde{\mathbf{T}} = \delta \mathbf{T} 
+ {\pounds}_{\delta x^{\mu}} \bar{\mathbf{T}} \,,
\end{equation}
where a tilde denotes quantities evaluated in the ``transformed'' coordinate system. 
The original and ``transformed'' - or new - coordinate systems are related by
\be
\label{coord_sys}
\tilde x^\mu=x^\mu+\delta x^{\mu}\,,
\ee
where $\delta x^{\mu}=[\delta t, \delta x^i]$ is the gauge generator. 
Throughout we use the Lie derivative, denoted by ${\pounds}_{\delta x^{\mu}}$.

\subsubsection{Matter and Metric Quantities}
\label{MatandMet}

From \eq{gauge_gen} and \eq{split_rho} we find the density
perturbation transforms simply as,
\begin{equation}
\label{denspertgtransEG}
\delta \tilde{\rho} = \delta \rho +  \dot{{\bar{\rho}}} \delta t
+  {\bar{\rho}}'  \delta x\,,
\end{equation}
as the background energy density depends on $t$ and $x$.
The perturbed 4-velocities, defined in \eq{4velpertEG}
transform as, 
\begin{equation}
\label{vu0Trans}
{\tilde{u}}^{\mu} = u^{\mu} - \dot{\delta x^{\mu}} \,,
\end{equation}
where $\mu = 0, 1, 2, 3$, or for each component,\\
\begin{equation}
\label{vutTrans}
{\tilde{u}}^0 = u^0 - \dot{\delta t} \,,
\end{equation}
\begin{equation}
\label{vuxTrans}
{\tilde{v}}^x = v^x - \dot{\delta x} \,,
\end{equation}
\begin{equation}
\label{vuyTrans}
{\tilde{v}}^y = v^y - \dot{\delta y} \,,
\end{equation}
\begin{equation}
\label{vuzTrans}
{\tilde{v}}^z = v^z - \dot{\delta z} \,,
\end{equation}
\\
We now transform the perturbed metric, using \eq{gauge_gen}, as
\begin{equation}
\label{metricgtransgen2}
{\delta \tilde{g}}_{\mu \nu} = \delta g_{\mu \nu} + \delta x^{\gamma} \partial_{\gamma}  {\bar{g}}_{\mu \nu} +  {\bar{g}}_{\gamma \nu} \partial_{\mu} \delta x^{\gamma} +  {\bar{g}}_{\mu \gamma} \partial_{\nu} \delta x^{\gamma} .
\end{equation}

The $0-0$ component of \eq{metricgtransgen2} allows us to find the transformation of the lapse function as,
\begin{equation}
\label{Phitrans}
\tilde{\Phi} = \Phi - \delta \dot{t} + \sqrt{\sigma} X \dot{\delta y}\,.
\end{equation}
From the transformation of the perturbations to the spatial trace part of the metric we find for the $x$ coordinate from \eq{metricgtransgen2},
\begin{equation}
\label{CxxTrans}
{\tilde{C}}_{xx} = C_{xx} + H \delta t + \delta x ' \,,
\end{equation}
for the $y$ coordinate,
\begin{equation}
\label{CyyTrans}
{\tilde{C}}_{yy} = C_{yy} + H_X \delta t  +  \frac{\sfx '}{\sfx} \delta x + \frac{\sqrt{\sigma}}{\kappa X} \partial_{y} \delta t + \partial_{y} \delta y \,,
\end{equation}
and for the $z$ coordinate,
\begin{equation}
\label{CzzTrans}
{\tilde{C}}_{zz} = C_{zz} + + H \delta t + \partial_{z} \delta z \,.
\end{equation}

The off diagonal spatial metric perturbations transform as,
\bea
\label{CxyTrans}
{\tilde{C}}_{xy} &=& C_{xy} + \frac{\sqrt{\sigma}}{a} \delta t' + \frac{\kappa X}{a} \delta y' + \frac{a}{X} \partial_y \delta x \,,\\
\label{CxzTrans}
{\tilde{C}}_{xz} &=& C_{xz} + \delta z' + \partial_z \delta x \,,\\
\label{CyzTrans}
{\tilde{C}}_{yz} &=& C_{yz} + \frac{a}{X} \partial_y \delta z + \frac{\sqrt{\sigma}}{a} \partial_z \delta t + \frac{\kappa X}{a} \partial_z \delta y  \,.
\eea
\\

For later use we here define a spatial metric perturbation, $\psi$, as,
\begin{equation}
\label{CurvatureEG1}
3 \psi = \frac{1}{2}\delta {g^k}_{k} = C_{xx} + C_{yy} + C_{zz}\,,
\end{equation}
i.e. the trace of the perturbed spatial metric, analogous to the curvature perturbation $\psi_\frw$ in perturbed FRW spacetime. As in our earlier work \cite{Leithes2014} the relation between $\psi$ here and the curvature perturbation in FRW can be most easily seen from the expansion scalar, given in \eq{ExpFacSph2}, which is
most similar to its FRW counterpart (see e.g.~\cite{MW2008},
Eq.~(3.19)).
\\
From the above  $\psi$ transforms as 
\begin{equation}
\label{psiTrans2}
3\tilde{\psi} = 3\psi + \left[2H + H_X \right] \delta t + \frac{X'}{X}  \delta x + \frac{\sqrt{\sigma}}{\kappa X} \partial_y \delta t + \partial_i \delta x^i  \,,
\end{equation}
where $i = x, y, z$.
\\

The mixed temporal-spatial perturbations of the metric, that is the
shift vector, from \eq{metricgtransgen2} transform as
\bea
\label{BxTrans}
{\tilde{B}}_x &=& B_x + a \dot{\delta x} - \frac{\delta t '}{a} + \frac{\sqrt{\sigma}X}{a} \delta y ' \,,\\
\label{ByTrans}
{\tilde{B}}_{y} &=& B_{y} + H_X \delta t + \frac{X'}{X} \delta x + \dot{\delta t} + \frac{\kappa X}{\sqrt{\sigma}} \dot{\delta y} - \frac{\partial_y \delta t}{\sqrt{\sigma} X} + \partial_y \delta y \,,\\
\label{BzTrans}
{\tilde{B}}_{z} &=& B_{z} + a \dot{\delta z} - \frac{\partial_z \delta t}{a} + \frac{\sqrt{\sigma}X}{a} \partial_z \delta y \,.
\eea

\subsubsection{Geometric Perturbed Quantities}
\label{Geom}

The expansion scalar, defined in \eq{ExpFacugen} with $u^\mu$ is calculated from the 4-velocity as given in \eq{4velpertEG} as,

\begin{equation}
\label{ExpFacSph2}
\Theta = 2H + 2H_X + 3\dot{\psi} + \dot{B}_y + \partial_i v^i
- 2H \Phi- 2H_X \Phi + 2 \frac{X'}{X} v^x \,.
\end{equation}
\\

In order to have the possibility to define hypersurfaces of
uniform expansion, where the perturbed expansion is zero, we require the transformation behaviour of the expansion scalar. We find $\Theta$ transforms as,
\begin{multline}
\begin{aligned}
\label{ExpFacSphNormTrans2}
&{\tilde{\Theta}}= \Theta + \partial_t \left[ (2H + 2H_X)\delta t + 2\frac{X'}{X} \delta x + \frac{\sqrt{\sigma}}{\kappa X} \partial_y \delta t + \partial_i \delta x^i + \dot{\delta t} + \frac{\kappa X}{\sqrt{\sigma}} \delta y - \frac{\partial_y \delta t}{\sqrt{\sigma} X} + \partial_y \delta y \right]\\& - \partial_i \delta x^i - [2H + 2H_X](\sqrt{\sigma}X\dot{\delta y} - \dot{\delta t}) \,.
\end{aligned}
\end{multline}

The transformation behaviour of $\Theta$ is rather complicated, and as such we do not use it to specify a gauge.

\subsection{Gauge Invariant Quantities}
\label{TheCurvePert}

We now use the results from the previous section to construct
gauge-invariant quantities. Fortunately we can use the results for the FRW background spacetime as guidance e.g. ~\cite{WMLL, Malik:2004tf}, where the evolution equation for the curvature perturbation on uniform density hypersurfaces, $\zeta$, is derived solely from the energy conservation equations, taking the large scales limit where spatial gradients become negligible.\\

\subsubsection{Gauge-invariant Quantities in EG Spacetime}
\label{EGST}

We now construct gauge-invariant quantities in the perturbed EG model, taking the FRW case as guidance.\\
Taking our earlier spatial metric perturbation, $\psi$, \eq{CurvatureEG1}, and its transformation behaviour, \eq{psiTrans2}, we can now choose a gauge condition, to get rid of the gauge artefacts, here $\delta t$. To this end, the uniform density gauge can then be specified by the choice $\wt{\delta\rho}\equiv 0$, which implies
\be
\label{delta_t_EG}
\delta t\Big|_{\delta \tilde{\rho}=0}= -\frac{1}{\dot{{\bar{\rho}}}}\left[\delta \rho
+  {\bar{\rho}}'  \delta x\right]\,.
\ee

Substituting this into \eq{psiTrans2}, the transformation of the trace of the perturbed spatial metric, we get,\\

\begin{equation}
\label{zetafull1}
- \SMTP = \psi 
- \frac{1}{3}\left[2 H + H_X \right] \left(\frac{\delta \rho + \bar{\rho} ' \delta x}{\dot{\bar{\rho}}} \right) 
+ \frac{X'}{3X} \delta x - \frac{\sqrt{\sigma}}{3\kappa X \dot{\bar{\rho}}}\left[\partial_y \delta \rho + \bar{\rho}'\partial_y \delta x\right] + \frac{1}{3} \partial_i \delta x^i \,,
\end{equation}
where we chose the sign convention and notation to coincide both with the
earlier work in FRW and our earlier work in LTB and Lema{\^\i}tre.\\

We now choose co-moving hypersurfaces to fix the remaining spatial
gauge freedoms. This gives for the spatial gauge generators from the
transformation of the 3-velocity perturbations as \eq{vuxTrans} and,
\be
\label{spatial_EG}
\delta x^i=\int v^i dt\,.
\ee
Substituting the above equations into \eq{zetafull1} we finally recover
the gauge-invariant spatial metric trace perturbation (SMTP) on co-moving, uniform density hypersurfaces,
\begin{equation}
\label{zetafull4}
- \SMTP = \psi - \frac{1}{3\dot{\bar{\rho}}}\left[2 H + H_X \right] \left(\delta \rho + \bar{\rho} ' \int v^x dt \right) 
+ \frac{X'}{3X} \int v^x dt - \frac{\sqrt{\sigma}}{3\kappa X\dot{\bar{\rho}}}\left[\partial_y \delta \rho + \bar{\rho}'\partial_y \int v^x dt\right] + \frac{1}{3} \partial_i \int v^i dt  \,.
\end{equation}\\

We can check by direct calculation, i.e.~by substituting \eq{psiTrans2}, \eq{denspertgtransEG}, and \eq{spatial_EG} into \eq{zetafull4}, that $\SMTP$ is gauge invariant.\\

Instead of using $\delta\rho$ to specify our temporal gauge, we could use the spatial metric trace perturbation, that is define hypersurfaces where $\wt\psi\equiv0$. This gives for $\delta t$
\be
\label{deltatfix}
\delta t \Big|_{\tilde{\psi}=0} = - \frac{1}{2H + H_X} \left[3 \psi + \left(\frac{\sfx '}{\sfx} \right) \delta x + \frac{\sqrt{\sigma}}{\kappa X} \partial_y \delta t \right].
\ee
which is a more complicated expression than that obtained from $\delta\rho$. However, in the large scale limit this would simplify to some degree. If we also assume co-moving relative to the background to eliminate extra spatial gauges we obtain the density perturbation on uniform spatial metric trace perturbation hypersurfaces,
\begin{equation}
\label{GIdenspertZeta}
\delta \tilde{\rho} \Big|_{\tilde{\psi}=0,\tilde{v^i}=0} = \delta \rho + \frac{\dot{\bar{\rho}}}{2H + H_X} \left[3 \psi \right] \,,
\end{equation}
which is gauge invariant under these conditions.
The density perturbation on uniform spatial metric trace perturbation
hypersurfaces \eq{GIdenspertZeta}, using \eq{denspertgtransEG} can be
written in terms of $\SMTP$, defined in \eq{zetafull4}, as,
\begin{multline}
\begin{aligned}
\label{GIdenspertZeta2}
&\delta \tilde{\rho} \Big|_{\tilde{\psi}=0,\tilde{v^i}=0} = \delta \rho +  \frac{\dot{\bar{\rho}}}{2H + H_X} \Big[ 3\SMTP - \left[2 H + H_X \right] \frac{1}{\dot{\bar{\rho}}} \left(\delta \rho + \bar{\rho} ' \int v^x dt \right) + \frac{X'}{X} \int v^x dt\\& - \frac{\sqrt{\sigma}}{\kappa X\dot{\bar{\rho}}}\left[\partial_y \delta \rho + \bar{\rho}'\partial_y \int v^x dt\right] + \partial_i \int v^i dt \Big] \,.
\end{aligned}
\end{multline}
This expression is more complicated than the equivalent in LTB \cite{Leithes2014} but as in this spacetime does allow us to relate the density perturbation at different times to the spatial metric trace perturbation, which, as we shall see in Section \ref{zeta_evol}, is conserved - or constant in time - on large scales for barotropic fluids.\\

\subsection{Evolution of $\SMTP$}
\label{zeta_evol}

Before deriving the evolution equation for spatial metric trace perturbation $\SMTP$, we briefly discuss the pressure perturbation in EG spacetime. We assume that the pressure $P=P(\rho,S)$, where $\rho$ is the density and $S$ the entropy of the system. We can expand the pressure and obtain the pressure perturbation as,
\be
\label{deltaP}
\delta P= \cs2\delta\rho+\dpn\,, 
\ee 
where $\dpn$ is the non-adiabatic pressure perturbation, and the adiabatic sound speed is defined as
\be
\label{cs2def}
\cs2\equiv \left.\frac{\p P}{\p\rho}\right|_S\,,
\ee
Since in EG background quantities are $t$ and $x$ dependent, we assume for now $\bar{P} \equiv \bar{P}(t,x)$, and find that
\be 
\label{cs2_tandx}
\cs2 = \frac{\dot{\bar{P}} + \bar{P}' v^x}{\dot{\bar{\rho}} +
  \bar{\rho}' v^x} \,. 
\ee

Since for our EG spacetime setup we initially assume pressureless dust as our matter content for simplicity, $\bar P=0$, we have that on uniform density
hypersurfaces $\delta P=\dpn$.
\\

The evolution equation for spatial metric trace perturbation on uniform density and co-moving hypersurfaces, $\SMTP$, using the time derivative of \eq{zetafull4}, \eq{PertEconsEG} and background conservation equation, \eq{HEGEngConsExpanded}, is
\begin{equation}
\label{zetaevol5}
\DOTSMTP = \frac{1}{3} \left( 2H + H_X +\frac{2\kappa}{\sigma} H_X \right) \frac{\dpn}{\bar{\rho}} \, .
\end{equation}
Therefore $\SMTP$ is conserved in the large scale limit for $\dpn=0$, e.g.~for barotropic fluids. This is similar to $\SMTP$ in Lema{\^\i}tre spacetime, but differs from LTB, which was conserved on all scales - at least for barotropic fluids \cite{Leithes2014}.\\
This result is also similar to the standard curvature perturbation, $\zeta$, in the FRW case \cite{WMLL}, although \eq{zetaevol5} is again slightly more complicated than the equivalent in FRW.

\section{A Perfect Fluid EG Spacetime}
\label{EGPerfect_sect}

Here we make a final extension to our work in EG cosmology to include a perfect fluid matter content. This could be useful when trying to examine the behaviour of perturbations - and thereby structure formation - in the early universe, when background pressure becomes significant.

\subsection{Perfect Fluid Background}
\label{EGPerfect_back}

The stress-energy-momentum tensor for a perfect fluid, \eq{Tmunu_up}, gives rise to the background energy conservation equation,

\begin{equation}
\label{HPerfectEGEngConsContracted}
\dot{\rho} + 2 H\rho + 3 H_X\rho + 2 H P + H_X P + \frac{2\kappa}{\sigma} H_X P = 0\,.
\end{equation}
%

\subsection{Perfect Fluid Perturbations}
\label{pert_EGPerfect}

As with the earlier pressureless dust model, \ref{pert_EG}, split quantities into a $t$ and $x$ dependent background part where necessary, and a perturbation depending on all four coordinates. We therefore decompose the pressure $P$ as follows,
\be
\label{split_P}
P=\bar{P}(t,x)+\delta P(x^\mu) \,.
\ee
The perturbed energy conservation equation is,
\begin{multline}
\begin{aligned}
\label{PertEconsPerfectEG}
& \delta \dot{\rho} + {\bar{\rho}}' v^x + \bar{\rho} {v^x}' + \bar{\rho} \partial_y v^y + \bar{\rho} \partial_z v^z - 2 \dot{\bar{\rho}} \Phi + 2 \dot{B}_y \bar{\rho} + \dot{C}_{xx} \bar{\rho} + \dot{C}_{yy} \bar{\rho} + \dot{C}_{zz} \bar{\rho} + 3 H_X \delta \rho + 2 H \delta \rho + 3 \frac{X'}{X} v^x \bar{\rho} + 2 \frac{\kappa \dot{X}}{\sqrt{\sigma}} v^y \bar{\rho} \\ & - 6 H_X \Phi \bar{\rho} - 4H \Phi \bar{\rho} + \frac{\partial_y \delta P}{\sqrt{\sigma} X} + \frac{\bar{\rho} \partial_y \Phi}{\sqrt{\sigma} X} + \frac{2 \kappa H_X \delta P}{\sigma} + 2H\delta P + H_X \delta P + \bar{P}' v^x + \bar{P} {v^{x}} ' + \bar{P} \partial_y v^y + \bar{P} \partial_z v^z + \frac{B_x}{a} {\bar{P}} '   
\\ & + \dot{C}_{xx} \bar{P} + \dot{C}_{yy} \bar{P} + \dot{C}_{zz} \bar{P} -4H\Phi \bar{P} -2H_X \Phi \bar{P} + 3\frac{X'}{X} v^x \bar{P} + \frac{3\bar{P}}{\sqrt{\sigma} X} \partial_y \Phi + \frac{B_x X' \bar{P}}{Xa} + \frac{2\bar{P}}{\sqrt{\sigma} X} \partial_y C_{yy} + \frac{C_{xy} X' \sqrt{\sigma} \bar{P}}{Xa}
\\ & + \frac{2\kappa \dot{X} v^y \bar{P}}{\sqrt{\sigma}} - \frac{4B_y\kappa H_X \bar{P}}{\sigma} + \frac{4C_{yy}\kappa H_X \bar{P}}{\sigma} + \frac{2 \dot{C}_{yy}\kappa \bar{P}}{\sigma} - \frac{2C_{xy}\kappa X' \bar{P}}{\sqrt{\sigma} X} - \frac{\bar{P} \sqrt{\sigma}}{\kappa X} \partial_y B_y + \frac{C_{xy} X' \bar{P}}{\sqrt{\sigma} aX} + \frac{{C_{xy}} ' \bar{P}}{\sqrt{\sigma} a} + \frac{\bar{P}}{\sqrt{\sigma} a} \partial_z C_{yz}
\\ & = 0 \,,
\end{aligned}
\end{multline}

\subsection{Gauge-invariant Quantities in Perfect Fluid EG Spacetime}
\label{PerfectEGST}

Our gauge-invariant spatial metric trace perturbation on co-moving, uniform density hypersurfaces, $\SMTP$, is constructed as in \ref{EGST} giving \eq{zetafull4} as before. Differences only become apparent when examining the evolution of $\SMTP$, due to the form of the perturbed energy conservation equation, \eq{PertEconsPerfectEG},  which now contains additional terms including the background pressure.
Also, since the density perturbation on uniform spatial metric trace perturbation hypersurfaces is derived simply from the transformation behaviour of the density perturbation to give \eq{GIdenspertZeta} the relation between it and $\SMTP$ remains \eq{GIdenspertZeta2}.

\subsection{Evolution of $\SMTP$ in a Perfect Fluid}
\label{zeta_evolPerfect}

If we take $\delta P = \dpn$ on uniform density hypersurfaces and use the large scale limit the evolution equation for $\SMTP$ is found using the time derivative of \eq{zetafull4}, the perturbed energy conservation equation for a perfect fluid, \eq{PertEconsPerfectEG}, and the background conservation equation for a perfect fluid, \eq{HPerfectEGEngConsContracted}, as
\begin{equation}
\label{zetaevol5Perfect}
\DOTSMTP = \frac{1}{3} \left( 2H + H_X +\frac{2\kappa}{\sigma} H_X \right) \frac{\dpn}{\bar{\rho} + \bar{P}} \, .
\end{equation}
which is conserved for barotropic fluids as with \eq{zetaevol5}.\\
We do not extend the work here to include a cosmological constant in our matter content since Dark Energy domination only occurs at relatively late times. By the epoch of Dark Energy domination it is likely rotation will have reduced to negligible levels and the spacetime will have effectively reduced to FRW. The evolution of perturbations and structure formation with Dark Energy has already been extensively researched using this cosmology.

\section{Discussion and conclusion}
\label{Conclusion}

In this paper we constructed gauge-invariant quantities in a perturbed EG spacetime. Specifically, we have constructed the
gauge-invariant spatial metric trace perturbation on co-moving, uniform density hypersurfaces, $\SMTP$.\\
We derived the evolution equations for $\SMTP$ and found that it is conserved on large scales for barotropic fluids (when $\dpn=0$). This is similar to the standard FRW result, where the similar gauge-invariant quantity, $\zeta$, is only conserved on large scales.\\

Deriving these results in EG is more involved than in either the LTB, Lema{\^\i}tre or FRW cases, in part because the background metric \eq{EGInterval} contains off-diagonal components complicating the form of background quantities and further complicating the perturbed quantities.\\

The main application of these results lies in providing conserved quantities in EG spacetime. Conserved quantities have proved to be very useful in standard cosmology with FRW, as they can be used to directly relate observable quantities, such as the density perturbation between earlier and later times, without having to solve the field equations. In addition, these conserved quantities can be used as a consistency check for numerical simulations, as for example those proposed in \cite{HUBBLETENSION}. By solving for the density contrast numerically and comparing this result to the one obtained by using $\SMTP$, the accuracy of the code can easily be checked.\\

As explained by Obukhov \cite{REFEG} the EG model is part of the wider class of Bianchi type III models, and identifying gauge-invariant conserved quantities within the class more broadly would be a useful avenue of further research, with a particular focus on spacetimes with rotation.\\

One attractive feature of the EG model is that it rapidly reduces to standard FRW with the ongoing expansion of the universe. Nevertheless, given the entire universe is initially rotating this should lead to observable consequences, not least of which would be identifying where this axis of rotation lies. This axis could lie within the currently observable universe, or beyond the particle horizon, but observational evidence should still exist in either case and further research would be needed to identify its location in either eventuality.\\
A rotating spacetime such as EG could also contribute to the anomalous CMB dipole (see e.g.~\citep{DipoleA,DipoleB}), particularly if we are situated a significant distance from the axis of rotation. Even given the magnitude of rotation has reached negligible levels in later times - and the cosmology currently conforms to FRW - this rotation in earlier epochs, and the location of its axis, could still have had a detectable effect on the CMB radiation. For an interesting review of some of the observations which are in tension with standard FRW cosmology see e.g.~\cite{Alurietal}.\\
Finally, recent detections by the JWST of large galaxies in the very early universe (see e.g.~\cite{EarlygalaxyA,EarlygalaxyB}) seem to disagree with current models of galaxy formation and the time-line for such formation. If in the very early universe the rotation still had significant dynamical effects on galaxy formation, as suggested by the galaxy rotation alignment results \cite{JWST}, perhaps there were further effects on the evolution of these early galaxies. Possibly with the entire spacetime rotating the entire universe shared some attributes - though on a far larger scale - with accretion disks around stars, possibly compressing the distribution of matter overall in the $z$-axis i.e. the axis of rotation as defined in the EG model \cite{REFEG}. This compression could aid and accelerate the formation of galaxies analogous to the formation of planets within stellar accretion disks. Again, this could prove a valuable avenue of further research.

\begin{acknowledgments}
The author is grateful to Tim Clifton, David Mulryne, Chris Clarkson and Juansher Jejelava for useful, insightful discussions and advice. AL is funded by The Roche School with grateful thanks to James Roche. The author is also grateful to Anton Petrov and Sabine Hossenfelder for inspiring a return to research.
\end{acknowledgments}



\appendix

\section{Additional Material for Background EG Spacetime}

In this section we present material that is not
necessary to follow the main body of this paper, but
might be useful for reproducing or extending some or all
of the calculations.

\subsection{The Non-zero Christoffel Symbols}
\label{Connections}
The non-zero connections are,
\begin{multline}
\begin{aligned}
\label{Christoffels}
&{\mathlarger{\mathlarger\Gamma}}^0_{00} = {\mathlarger{\mathlarger\Gamma}}^2_{02} = {\mathlarger{\mathlarger\Gamma}}^2_{20} = H_{\sfx} \,, 2 {\mathlarger{\mathlarger\Gamma}}^0_{01} = {\mathlarger{\mathlarger\Gamma}}^0_{10} = \frac{2}{3} {\mathlarger{\mathlarger\Gamma}}^2_{12} = \frac{2}{3} {\mathlarger{\mathlarger\Gamma}}^2_{21} = \frac{X'}{X} \,,  {\mathlarger{\mathlarger\Gamma}}^0_{02} = \frac{1}{2} {\mathlarger{\mathlarger\Gamma}}^0_{20} = - {\mathlarger{\mathlarger\Gamma}}^2_{22} =\frac{\kappa \dot{X}}{\sqrt{\sigma}} \,, {\mathlarger{\mathlarger\Gamma}}^0_{11} = {\mathlarger{\mathlarger\Gamma}}^0_{33} = a \dot{a} \,, \\ &{\mathlarger{\mathlarger\Gamma}}^0_{12} = {\mathlarger{\mathlarger\Gamma}}^0_{21} = \frac{\kappa X'}{\sqrt{\sigma}} - \frac{\sqrt{\sigma} X'}{2} \,, {\mathlarger{\mathlarger\Gamma}}^0_{22} = \kappa X \dot{X} \,, {\mathlarger{\mathlarger\Gamma}}^1_{01} = {\mathlarger{\mathlarger\Gamma}}^1_{10} =  {\mathlarger{\mathlarger\Gamma}}^3_{03} = {\mathlarger{\mathlarger\Gamma}}^3_{30} = H \,, {\mathlarger{\mathlarger\Gamma}}^1_{02} = {\mathlarger{\mathlarger\Gamma}}^1_{20} = - \frac{\sqrt{\sigma} X'}{2 a^2} \,, {\mathlarger{\mathlarger\Gamma}}^1_{22} = - \frac{\kappa X X'}{a^2} \,, \\ &{\mathlarger{\mathlarger\Gamma}}^2_{00} = \frac{\sqrt{\sigma} \dot{X}}{\kappa X^2} \,, {\mathlarger{\mathlarger\Gamma}}^2_{01} = {\mathlarger{\mathlarger\Gamma}}^2_{10} = \frac{\sqrt{\sigma} X'}{2\kappa X^2} \,, {\mathlarger{\mathlarger\Gamma}}^2_{11} = - \frac{a \dot{a}}{\sqrt{\sigma} X} \,, 
\end{aligned}
\end{multline}

\subsection{The Ricci Tensor Components}
\label{RicciTensorCompenents}

The $0-0$ component of the Ricci Tensor is,
\begin{equation}
\label{RicciTensor00}
R_{00} = - \frac{\ddot{X}}{X} - 2{H_{\sfx}}^2 - 2\frac{\ddot{a}}{a} + 2 H H_{\sfx} + \frac{\sigma (X')^2}{2 \kappa X^2 a^2} \,.
\end{equation}

The $0-1$ component is,
\begin{equation}
\label{RicciTensor01}
R_{01} = - \frac{3\dot{X}'}{2X} - H_{\sfx} \frac{X'}{X} + 2 H \frac{X'}{X} + \frac{\sigma H_X X'}{2 \kappa X} \,.
\end{equation}

The $0-2$ component is,
\begin{equation}
\label{RicciTensor02}
R_{02} = \frac{\kappa \ddot{X}}{\sqrt{\sigma}} - \frac{\sqrt{\sigma} X''}{2 a^2} + \frac{\kappa H_X \dot{X}}{\sqrt{\sigma}} + \frac{\sqrt{\sigma} (X')^2 }{2 X a^2} + \frac{2 \kappa H \dot{X}}{\sqrt{\sigma}} - \sqrt{\sigma} H_X \dot{X} \,.
\end{equation}

The $1-1$ component is,
\begin{equation}
\label{RicciTensor11}
R_{11} = \dot{a}^2 + a \ddot{a} - \frac{5X''}{2X} - \frac{5(X')^2}{4X^2} + 2a\dot{a}H_X + \frac{\sigma (X')^2}{2 \kappa X^2} \,.
\end{equation}

The $1-2$ component is,
\begin{equation}
\label{RicciTensor12}
R_{12} = \frac{\kappa \dot{X}'}{\sqrt{\sigma}} - \frac{\sqrt{\sigma}\dot{X}' }{2} + \frac{3\kappa H_X X'}{2 \sqrt{\sigma}} - \sqrt{\sigma} H_X X' \,.
\end{equation}

The $2-2$ component is,
\begin{equation}
\label{RicciTensor22}
R_{22} = \kappa \dot{X}^2 + \kappa X \ddot{X} + \frac{(\kappa X')^2}{a^2} - \frac{\kappa X X''}{a^2} - \frac{3 \kappa^2 \dot{X}^2}{\sigma} + 2\kappa X \dot{X}H - \frac{\sigma (X')^2}{2 a*2} \,.
\end{equation}

The $3-3$ component is,
\begin{equation}
\label{RicciTensor33}
R_{33} = \dot{a}^2 + a\ddot{a} + 2 a\dot{a}H_X \,.
\end{equation}
The $0-3$, $1-3$ and $2-3$ components are all zero.
\\

\subsection{The Ricci Scalar}
\label{RicciScalarSec}

From the Ricci Tensor we construct the Ricci Scalar, $R$,
\begin{equation}
\label{RicciScalar}
R = 4 \frac{\ddot{a}}{a} + 2 \frac{\ddot{X}}{X} + 2 \frac{\kappa \ddot{X}}{\sigma X} - \frac{9 X''}{2X a^2} + {H_X}^2 + 4 H H_X - \frac{\sigma (X')^2}{2\kappa X^2 a^2} - \frac{\kappa {H_X}^2}{\sigma} + \frac{3(X')^2}{4X^2 a^2} + \frac{4\kappa H H_X}{\sigma} + 2 H^2 \,.
\end{equation}

\subsection{The Einstein Field Equations}
\label{EFEs}

For the $0-0$ component we find,

\begin{equation}
\label{HEinstein00}
H^2 - \frac{3{H_X}^2}{2} + 4 H H_X + \frac{\sigma (X')^2}{4\kappa X^2 a^2} + \frac{\kappa \ddot{X}}{\sigma X} - \frac{9X''}{4X a^2} - \frac{\kappa {H_X}^2}{2\sigma} + \frac{3 (X')^2}{8 X^2 a^2} + \frac{2\kappa H H_X}{\sigma} = 8 \pi G \rho  \,.
\end{equation}
For the $0-1$ component we find,
\begin{equation}
\label{HEinstein01}
\frac{H_X X'}{X} + \frac{3 \dot{X}'}{2X} - \frac{2H X'}{X} - \frac{\sigma H_X X'}{2\kappa X} = 0  \,.
\end{equation}
For the $0-2$ component we find,
\begin{equation}
\label{HEinstein02}
\frac{7\sqrt{\sigma}X''}{4a^2} + \frac{3\kappa H_X \dot{X}}{2\sqrt{\sigma}} + \frac{\sigma^{\frac{3}{2}} (X')^2}{4\kappa X a^2} + \frac{\sqrt{\sigma} (X')^2 }{8 X a^2} - \frac{3\sqrt{\sigma} H_X \dot{X}}{2} - \sqrt{\sigma} \ddot{X} - \frac{2\sqrt{\sigma} X \ddot{a}}{a} - 2\sqrt{\sigma}H\dot{X} - \sqrt{\sigma}H^2 X = 0  \,.
\end{equation}
For the $1-1$ component we find,
\begin{equation}
\label{HEinstein11}
a\ddot{a} + \frac{X''}{4X} + \frac{13 (X')^2}{8 X^2} + \frac{\ddot{X}a^2}{X} + \frac{{H_X}^2a^2}{2} + \frac{\kappa \ddot{X}a^2}{\sigma X} + \frac{2\kappa a \dot{a} H_X}{\sigma} - \frac{3 \sigma (X')^2}{4\kappa X^2} - \frac{\kappa {H_X}^2 a^2}{2\sigma} = 0  \,.
\end{equation}
For the $1-2$ component we find,
\begin{equation}
\label{HEinstein12}
\frac{\kappa \dot{X}'}{\sqrt{\sigma}} - \frac{\sqrt{\sigma}\dot{X}' }{2} + \frac{3\kappa H_X X'}{2 \sqrt{\sigma}} - \sqrt{\sigma} H_X X' = 0  \,.
\end{equation}
For the $2-2$ component we find,
\begin{equation}
\label{HEinstein22}
\frac{\kappa \dot{X}^2}{2} + \frac{5\kappa (X')^2}{8a^2} + \frac{5\kappa XX''}{4a^2} - \frac{5\kappa^2 \dot{X}^2}{2\sigma} - \frac{2\kappa \ddot{a} X^2}{a} - \frac{\sigma (X')^2}{4a^2} - \frac{\kappa^2 \ddot{X}X}{\sigma} - \frac{2\kappa^2 X\dot{X}H}{\sigma} - \frac{\kappa X^2 H}{a} = 0  \,.
\end{equation}
For the $3-3$ component we find,
\begin{equation}
\label{HEinstein33}
a\ddot{a} + \frac{a^2 \ddot{X}}{X} + \frac{a^2 {H_X}^2}{2} + \frac{\kappa\ddot{X}a^2}{\sigma X} + \frac{3(X')^2}{8X^2} + \frac{2\kappa a\dot{a} H_X}{\sigma} - \frac{\sigma(X')^2}{4\kappa X^2} - \frac{9X''}{4X} - \frac{\kappa a^2 {H_X}^2}{2\sigma} = 0  \,.
\end{equation}

The $0-3$, $1-3$ and $2-3$ components are all identically zero.

\section{Additional Material for Perturbed EG Spacetime}
\label{PerturbedEGSec}

Using the constraint \eq{MetricConstraint} we get the contravariant perturbed
metric components,
\begin{equation}
\label{EGMetricperturbationsuppy}
\delta g^{\mu \nu}=\begin{pmatrix}
  2\Phi & \frac{B_x}{a} & - \frac{B_y}{\sqrt{\sigma} \sfx} & \frac{B_z}{a} \\
\\  
   \frac{B_x}{a} & -2 \frac{C_{xx}}{a^2} & - \frac{C_{xy}}{a X} & - \frac{C_{xz}}{a^2} \\
\\   
  - \frac{B_y}{\sqrt{\sigma} \sfx} & - \frac{C_{xy}}{a X} & -2 \frac{ C_{yy}}{\kappa\sfx^2} & - \frac{C_{yz}}{a X} \\
\\  
  \frac{B_z}{a} & - \frac{C_{xz}}{a^2} & - \frac{C_{yz}}{a X} &  -2 \frac{C_{zz}}{a^2}
\end{pmatrix} \,.
\end{equation}

%






\end{document}